\documentclass[prb,twocolumn,showpacs,superscriptaddress,floatfix]{revtex4-1}

\usepackage{graphicx}
\usepackage{verbatim}
\usepackage{comment}
\usepackage{amssymb}
\usepackage{amsmath}
\usepackage{stmaryrd}
\usepackage{color}

\newcommand{\rb}{Rb$_2$Cu$_3$SnF$_{12}$}

\begin{document}

\def\l{{\lambda}}
\def\rb{{Rb$_2$Cu$_3$SnF$_{12}$}}

\title{Thermodynamics and phase transitions for the Heisenberg model on the
pinwheel distorted kagome lattice}
\author{Ehsan Khatami}
\affiliation{Department of Physics, Georgetown University, Washington, D.C.
20057, USA}
\author{Rajiv R. P. Singh}
\affiliation{Physics Department, University of California, Davis, California
95616, USA}
\author{Marcos Rigol}
\affiliation{Department of Physics, Georgetown University, Washington, D.C.
20057, USA}

\begin{abstract}
We study the Heisenberg model on the pinwheel distorted kagome lattice as
observed in the material Rb$_2$Cu$_3$SnF$_{12}$. Experimentally relevant
thermodynamic properties at finite temperatures are computed utilizing numerical
linked-cluster expansions. We also develop a Lanczos-based zero-temperature
numerical linked-cluster expansion to study the approach of the pinwheel
distorted lattice to the uniform kagome-lattice Heisenberg model. We find 
strong evidence for a phase transition before the uniform limit is reached,
implying that the ground state of the kagome-lattice Heisenberg model is likely
not pinwheel dimerized and is stable to finite pinwheel-dimerizing
perturbations.
\end{abstract}

\maketitle


\section{Introduction}

Among all of the realistic quantum spin models studied to date, the spin-half
kagome-lattice Heisenberg model (KLHM) stands out as one with perhaps the most
esoteric low-energy properties.\cite{lhuillier-review} The classical Heisenberg
model on this lattice of corner-sharing triangles has an exponentially large
number of ground states. Fluctuations lead to a selection of $\sqrt{3}\times
\sqrt{3}$ order.\cite{huse-rutenberg} However, such an order is clearly absent
in the spin-half model.\cite{singh-huse1} A comparison between exact
diagonalization (ED) results in different lattice geometries highlights the
unusual spin physics of this model.\cite{lhuillier-review} Square and triangular
lattice models show a tower of rotor states separated from the rest of the
spectrum, as expected for a system with spontaneously broken spin-rotational
symmetry. In contrast, the KLHM shows that (i) the energy gaps are an order of
magnitude smaller than on the other lattices, and (ii) there are a very large
number of low-lying singlet states. These data have been interpreted to support
the existence of (i) a very small or even zero triplet gap,\cite{sindzingre} 
(ii) gapless singlet excitations,\cite{misguich} and (iii) an exponentially
large number of singlets below the lowest triplet.\cite{mila}

Numerical and analytical studies have suggested alternative ground states, which
include several different valence-bond crystals (VBCs), as well as gapped and 
gapless resonating valence-bond (RVB) spin
liquids.\cite{c_zeng_90,mila,r_singh_07,b_yang_09,y_simeng_11} 
Most numerical studies, including the recent density matrix renormalization
group (DMRG) study of Yan {\em et al.},\cite{y_simeng_11} suggest a singlet
ground state with a gap to spin excitations in the thermodynamic limit. While
earlier numerical studies\cite{lhuillier-review} provided evidence for a spin
gap of the order of $J/20$, where $J$ is the nearest-neighbor exchange constant,
and a much smaller gap to singlet excitations (possibly even gapless), recent
DMRG work puts the singlet gap at $0.05 J$, with a larger gap for the triplets.
These studies should be contrasted with the experimental results for the
KLHM.\cite{He3,herbertsmithites,volborthites} Among them, the herbertsmithites
with structurally perfect kagome planes have attracted the most interest. The
experimental data for these systems show no evidence of a spin gap. However,
substitutional impurities and Dzyaloshinskii-Moria anisotropy are present in
these systems and may be playing a role in producing the eventual thermodynamic
phase.\cite{M_rigol_07c,DM-phase-transition}

\begin{figure}[!b]
\centerline {\includegraphics*[width=0.47\textwidth]{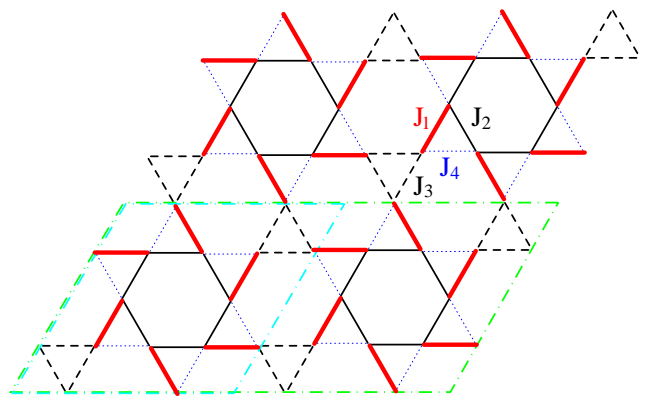}} 
\vspace{-0.2cm}
\caption{(Color online) Kagome lattice with bond anisotropy as relevant to the
\rb~material. The strength of the bonds ($J_{\alpha}$) can be inferred by their
thickness. Red (thick) bonds show dimers of the pinwheel VBC order with a
$12$-site unit cell, which is the ground state when $J_1$ is sufficiently larger
than the other exchange interactions. The dotted-dashed lines show the $12$-site
and $24$-site periodic clusters used in ED.}
\label{fig:lattice}
\end{figure}

One kagome material that does show gapped spin excitations is
\rb.\cite{k_morita_08,k_matan_10} This material also has substantial pinwheel
distortions, which results in a 12-site unit cell\cite{b_yang_09} and four
different nearest-neighbor exchange constants, with the ratio of the largest to 
smallest being about two (see Fig.~\ref{fig:lattice}). Given the many
theoretical proposals for spin-gapped and VBC phases for the KLHM and this
experimental finding, it is natural to ask if the uniform KLHM has a tendency to
spontaneously dimerize in this pattern. 

Here, we study this possibility and the role such a pattern of distortion might 
play in the ground state of the uniform KLHM. In earlier series-expansion
studies, different dimer configurations were studied order by order in powers
of a parameter $\lambda$, which takes one from a dimerized to a uniform 
Hamiltonian.\cite{r_singh_07} In each successive order, the set of dimer states 
was reduced by keeping only those that were lowest in energy. In low powers of
$\lambda$, the pinwheel-dimer state has one of the highest energies, as it has
no resonating hexagons, and was not considered further.\cite{r_singh_07}
However, all dimer states were found to be degenerate to the second order in
perturbation theory and stayed within a few percent of each other for
$\lambda=1$ to the highest order studied. Thus, it is possible that higher
powers of $\lambda$ can stabilize a different state than the ones picked out
perturbatively. Standard series expansions in $\lambda$ are clearly not suitable
to study such a selection.

We develop here a numerical linked-cluster expansion (NLCE)\cite{m_rigol_06} 
to study this problem. This is a graphical method based on the principle of
inclusion and exclusion for calculating the properties of an infinite system by
summing over contributions from all smaller clusters. Its convergence is not
based on the existence of a small parameter, but rather on having a short
correlation length, so that contributions from larger clusters become small.
Such an expansion always converges at high temperatures, but can also converge
at low temperatures for models where the correlation length remains
short.\cite{m_rigol_06}

A suitably chosen NLCE scheme can also converge at $T=0$ if the system has
long-range VBC order, but the fluctuating correlation length on top of the VBC
order remains small. By building the clusters in terms of the dimers of the
pinwheel VBC order (see Appendix for details of the expansion), the properties
of such a phase can be calculated. There is no small parameter in the expansion,
as the properties of the finite clusters are calculated exactly. For
concreteness, we introduce a family of models that depend on a parameter $\l$,
defined in the same way as for the series expansion. As long as our NLCE shows
internal convergence, one can study any ground-state property. The failure to
converge signals a phase transition away from the VBC phase.

We first use the NLCE to study the specific heat ($C_v$), entropy ($S$), and 
spin susceptibility ($\chi$) of the pinwheel distorted kagome lattice with 
exchange constants that describe
Rb$_2$Cu$_3$SnF$_{12}$.\cite{k_morita_08,k_matan_10} Our results are accurate in
the thermodynamic limit and converge at almost all temperatures (including
$T=0$). They also compare very well with the experimental data for the uniform
susceptibility, confirming the pinwheel valence-bond nature of the ground state
of this material. We then consider a model in which the exchange constants are
unity for the strong bonds in the pinwheel order and $\l$ for all others. We
study the thermodynamics of this model for $\l$ from the dimerized ($\l \ll 1$)
to the uniform KLHM ($\l=1$). Since, sufficiently away from $\l=1$, the NLCE
converges at $T=0$, we use a Lanczos-based NLCE and include larger cluster sizes
to study ground-state properties (see Appendix for details). We find that a VBC
order parameter decreases as $\l$ increases and, according to a power-law fit,
vanishes at $\l_c<1$. The results for finite-size clusters with periodic
boundary conditions point to a first-order phase transition from the VBC phase
to a spin-liquid-like phase at $\l_c\gtrsim 0.9$. 


\section{The model}

The KLHM Hamiltonian is written as:
\begin{equation}
\hat{H}=\sum_{\left<ij\right>_{\alpha}}J_{\alpha}\, {\bf \hat{S}}_i \cdot {\bf
\hat{S}}_j
\end{equation}
where $\left<..\right>$ denotes nearest neighbors, ${\bf \hat{S}}_i$ is the
spin-$1/2$ 
vector on site $i$, and $\alpha$ is the bond type. We consider the bond
anisotropy seen 
in Fig.~\ref{fig:lattice} with four distinct $\alpha$'s.


\section{Results}

\subsection{Thermodynamics of Rb$_2$Cu$_3$SnF$_{12}$}

\begin{figure}[t]
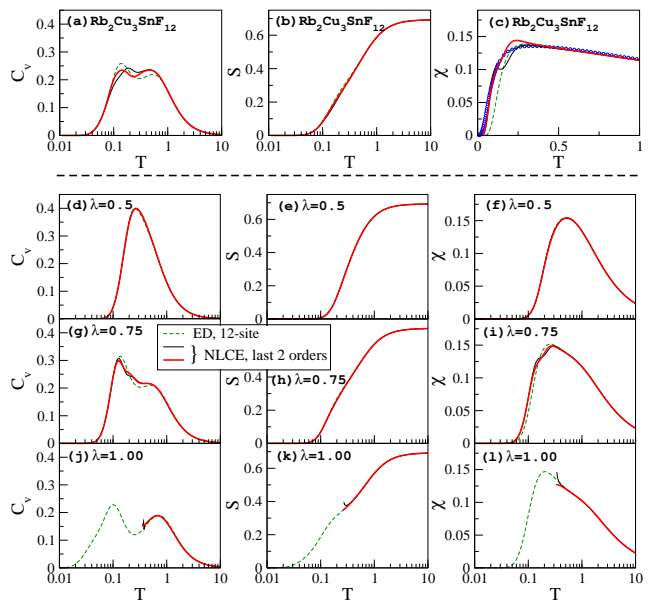

\centerline
{\includegraphics*[width=0.47\textwidth]{Theor_FiniteT_5.eps}} 
\centerline
{\includegraphics*[width=0.47\textwidth]{Theor_FiniteT_3.eps}} 
\vspace{-0.2cm}
\caption{(Color online) Specific heat, entropy, and uniform spin susceptibility
per site for the pinwheel distorted kagome lattice of Fig.~\ref{fig:lattice}
with (a)-(c) $J_2=0.95J_1$, $J_3=0.85J_1$, and $J_4=0.55J_1$ (the model for
\rb), and (d)-(l) $J_2=J_3=J_4=\lambda J_1$. The last two orders of the bare
sums in the NLCE are shown. They are $5^{\text{th}}$ and $6^{\text{th}}$ orders
for $\lambda=0.50$, and $8^{\text{th}}$ and $9^{\text{th}}$ orders for
$\lambda=0.75$ and $\lambda=1.00$, and in (a)-(c). Dashed lines are results from
ED for the periodic $12$-site cluster. Blue circles in (c) are experimental
results from Ref.~\onlinecite{k_morita_08} (see text).}
\label{fig:lambda}
\end{figure}

We study the thermodynamics of the anisotropic KLHM shown in
Fig.~\ref{fig:lattice}, with the
exchange constants suggested for \rb~in a recent experiment.\cite{k_matan_10} 
We take $J_2=0.95J_1$, $J_3=0.85J_1$, and $J_4=0.55J_1$, and set the unit of
energy to $J_1$. In Figs.~\ref{fig:lambda}(a)-\ref{fig:lambda}(c), we show
results for the $8^{\text{th}}$ and $9^{\text{th}}$ orders of bare NLCE
sums,\cite{m_rigol_06} along with results from ED on a $12$-site periodic
cluster, for $C_v$, $S$, and $\chi$ per site. In addition to the high-$T$
region, the thermodynamic quantities also fully converge in a low-$T$ window
between $0$ and $0.1$. This shows that clusters in our dimer expansion around
the pinwheel VBC order are able to capture the low-$T$ properties very well,
which in turn, supports the finding that this ordered phase describes the ground
state of \rb.\cite{b_yang_09,k_matan_10} 

In Fig.~\ref{fig:lambda}(a), note the relatively flat $C_v$ of this model for
$T$ between $0.1$ and $1.0$ [Fig.~\ref{fig:lambda}(a)], reminiscent of a
classical gas. The entropy in that region varies almost linearly with $\log{T}$
[Fig.~\ref{fig:lambda}(b)] . The ED results for $C_v$ exhibit a double-peak
structure but clearly underestimate the value of the high-$T$ peak as seen in
Fig.~\ref{fig:lambda}(a). Large finite-size effects in ED can also be seen at
low $T$ in Fig.~\ref{fig:lambda}(c).

Figure~\ref{fig:lambda}(c) shows a comparison between the experimentally
measured\cite{k_morita_08} $\chi$ and the NLCE results using $J_1=234$ K and the
$g$ factor, $g=2.46$ (needed to match the two at high $T$). We find a remarkable
agreement between the NLCE results within its convergence region ($T> 0.4$ and
$0<T<0.15$) and the experiments [Fig.~\ref{fig:lambda}(c)]. The fact that the
zero-$T$ fit of the spectra in Ref.~\onlinecite{k_matan_10} and the finite-$T$
susceptibility correspond to roughly the same parameters also shows that the
exchange constants are nearly temperature independent. Small deviations at
$T<0.07$ can be attributed to the uncertainty in the impurity
subtraction.\cite{k_morita_08}


\subsection{Thermodynamics of the model with $J_2=J_3=J_4=\lambda J_1$}

We are interested in the behavior of thermodynamic quantities and the NLCE
convergence at low $T$ for different $\l$ values.\cite{Note1} In
Figs.~\ref{fig:lambda}(d)-\ref{fig:lambda}(l), we show $C_v$, $S$, and $\chi$
for $\l=0.50$, $0.75$, and $1.00$.
Figures~\ref{fig:lambda}(d)-\ref{fig:lambda}(f) show excellent convergence of
the NLCE results at all $T$ in the $5^{\text{th}}$ and $6^{\text{th}}$ orders
for $\l=0.5$. Note that the specific heat in Fig.~\ref{fig:lambda}(d) has a
single peak at $T\sim0.27$ with no other features, which is characteristic of a
single dimer. 

Upon increasing $\l$ to $0.75$, the NLCE results still converge at nearly all
temperatures by increasing the order to $9$. Despite this convergence, which
suggests that the physics of the VBC is still dominant, the temperature
dependence of the specific heat, entropy, and susceptibility 
changes from those of $\l=0.5$. The peak in the specific heat splits into two 
peaks [Fig.~\ref{fig:lambda}(g)] and the entropy develops a hump in that region 
[Fig.~\ref{fig:lambda}(h)]. Also, the peak in susceptibility broadens and moves
to lower $T$ [Fig.~\ref{fig:lambda}(i)]. Unlike the $\l=0.5$ case, due to
finite-size effects, the $12$-site ED results do not completely agree with the
NLCE results when $\l=0.75$, and again, underestimate the specific heat for
$0.2\lesssim T \lesssim 0.5$.

The convergence of the NLCE results changes completely in the uniform limit
($\l=1$), as depicted in Figs.~\ref{fig:lambda}(j)-\ref{fig:lambda}(l). Similar
to previous NLCE calculations that use the site or triangular expansions on the
kagome lattice,\cite{m_rigol_06} the series do not converge below $T\sim0.3$.
Hence, the low-temperature physics can no longer be described in the dimerized
picture.


\subsection{The ground state}

To study ground-state properties directly, we use the Lanczos algorithm and
carry out the NLCEs to the 13th order. We calculate the bond energies
$B_{\alpha}$ per bond of each type in Fig.~\ref{fig:lattice}. The difference
between $B_1$ and any other bond energy, {\em e.g.}, $B_2-B_1$, can be used as
the order parameter. For small $\l$, when the system is almost completely
dimerized in the pinwheel pattern, $B_1$ is dominant and the order parameter is
maximum. As $\l$ increases, all $B_{\alpha}$ should approach the same value,
which at $\l=1$ is half the ground-state energy per site. Although this order
parameter can never be exactly zero for any $\l \neq 1$, there is the
possibility of a phase transition from the bond-solid order to some other phase
for $\l_c <1$, beyond which the difference in bond energies arises from the
difference in the exchange constants, not dimerization. Naively, the order
parameter would vanish linearly with $\l$ in this region. 

\begin{figure}[t]
\centerline {\includegraphics*[width=0.47\textwidth]{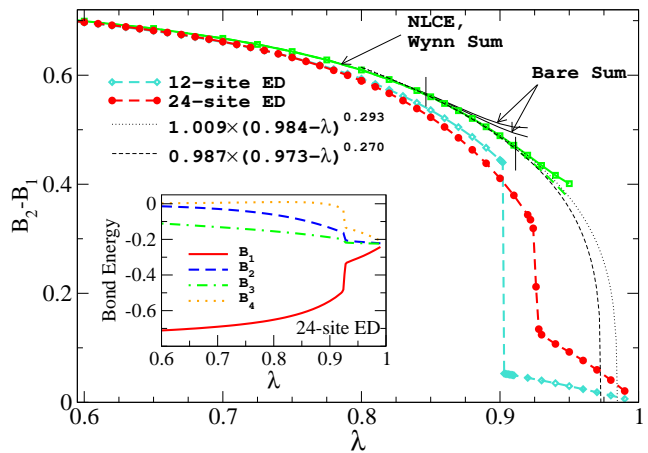}} 
\vspace{-0.2cm}
\caption{(Color online) VBC order parameter at $T=0$ vs the anisotropy
parameter $\lambda$ for $J_2=J_3=J_4=\lambda J_1$. The results for finite-size
clusters show a first-order transition. Bare NLCE sums do not converge beyond
$\lambda\sim 0.85$. We use the Wynn sum with four cycles of
improvement\cite{m_rigol_06} to extend the convergence to $\lambda \sim 0.9$. 
Results for the last two orders of the former are fit to
$A(\lambda_c-\lambda)^B$ (thin dotted and dashed lines). Vertical solid lines
show the region used for the fit. The inset shows the four bond energies for the
24-site ED vs $\lambda$.}
\label{fig:B1-B2}
\end{figure}

In Fig.~\ref{fig:B1-B2}, we show the dependence of $B_2-B_1$ on $\l$. We find
very good convergence for small values of $\l$. However, for $\l>0.85$, the bare
sums in the NLCE do not converge. Re-summing the series using the Wynn
algorithm\cite{m_rigol_06} extends the region of convergence to $\l\sim 0.91$
(last two orders are shown in Fig.~\ref{fig:B1-B2}). This is, in fact, an
indication that a small pinwheel distortion of the kagome lattice fails to 
stabilize the VBC order and that there is a transition at $0.9<\l_c<1$ to
another phase [presumably a spin-liquid (SL) phase] in the thermodynamic limit.
In Fig.~\ref{fig:B1-B2}, we also show the ED results from the $12$- and
$24$-site clusters for the entire region of $\l$. Both clusters exhibit a clear
first-order transition with $\l_c<1$. In the inset of Fig.~\ref{fig:B1-B2}, we
show the behavior of individual bond energies for the $24$-site cluster. It is
interesting to see that the bond energy $B_3$ of the so-called ``empty
triangles''\cite{r_singh_07} is the least affected by the transition. We should
stress that $B_2-B_1$ in the ED results exhibit large finite-size effects in the
region where the NLCE is converged, and increasing the cluster size from 12
to 24 does not improve the value of $B_2-B_1$ for $\l<\l_c$. As one can see
in Fig.~\ref{fig:B1-B2}, for $\l<\l_c$, the results from the $24$-site cluster
are further away from the (thermodynamic limit) NLCE results than those from the
$12$-site cluster.

A power-law fit of the high-$\l$ NLCE results for the order parameter allows us
to estimate $\l_c$ in the event of a second-order phase transition. We find
$\l_c \sim 0.98$ when considering the last two orders of the NLCE after 
resummation. Results for the fit are shown as thin dotted and dashed lines in
Fig.~\ref{fig:B1-B2}. If the transition is actually first order, $\l_c$ will be
smaller. 

\begin{figure}[t]
\centerline
{\includegraphics*[width=0.47\textwidth]{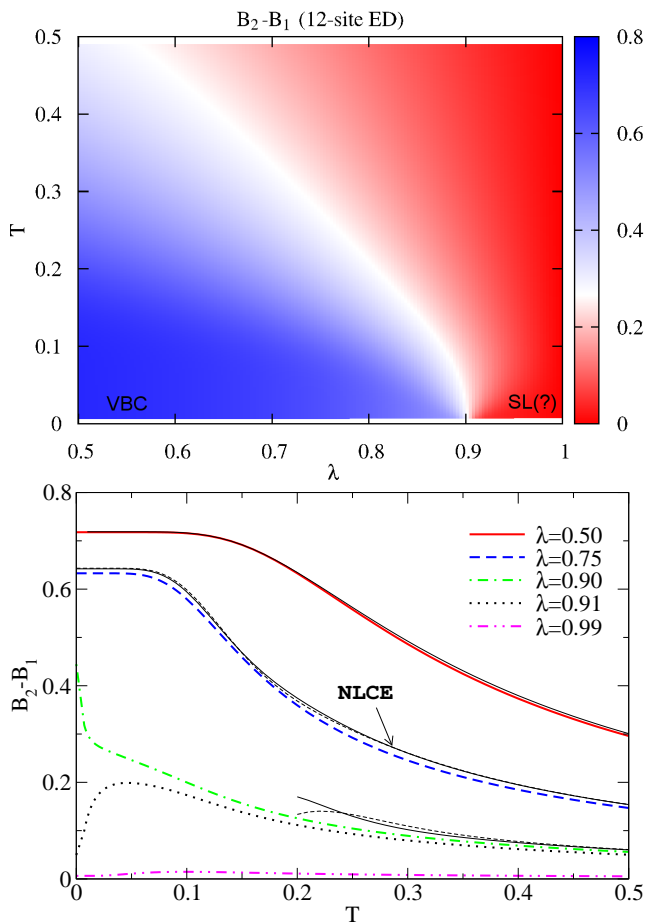}} 
\centerline {\includegraphics*[width=0.47\textwidth]{FiniteT_order.eps}}
\vspace{-0.2cm}
\caption{(Color online) Upper panel: Plot of the VBC order parameter $B_2-B_1$
in the $T$-$\lambda$ plane for the 12-site ED calculations. The decrease in
$B_2-B_1$ by increasing $\lambda$ from the left VBC region to the right region
is less abrupt at higher temperatures. Lower panel: The same quantity vs $T$ for
five different values of $\lambda$ below and above the transition point. Thin
lines are the $7^{\text{th}}$ and $8^{\text{th}}$ order NLCE, and thick lines
are the results for the 12-site cluster.}
\label{fig:12_3D}
\end{figure}

\subsection{Order parameter at finite temperatures}

Finally, in connection with the experiments, and to explore the effects of a
possible quantum critical region, we use NLCEs and ED on the $12$-site cluster 
to examine the behavior of $B_2-B_1$ at finite $T$. The contour plot in the
upper panel of Fig.~\ref{fig:12_3D} shows the order parameter in the $T-\l$
plane obtained from the ED of the 12-site cluster. The blue area (dark area in
the left) for $\l\lesssim0.9$ at low temperatures indicates the VBC phase, and
the red area (dark area in the right) for $\l\gtrsim0.9$ belongs to the other
phase. Separating these two phases is a crossover region, shown by the light
color in that figure, which starts at the finite-size transition point of the
12-site cluster ($\l_c\sim 0.91$) and is broadened as it moves to smaller $\l$
at higher $T$. We expect the same qualitative behavior in the thermodynamic
limit, with a transition point that is closer to the $\l=1$ axis.

To compare with the NLCE results, in the lower panel of Fig.~\ref{fig:12_3D}, we
plot the order parameter for different values of $\l$ below and above the
transition point. An interesting feature there is the initial low-temperature
rise of the order parameter above but very close to $\l_c$ as $T$ is increased
from zero. In this region, thermal fluctuations initially increase the value of
the order parameter before suppressing it at higher $T$. If the transition in
the thermodynamic limit is second order or weakly first order and closer to
$\lambda=1$, there could be a quantum critical region that controls the
temperature dependence of the KLHM over some temperature
range.\cite{M_rigol_07c}

\section{Summary}

In summary, we have studied the KLHM on a pinwheel distorted kagome lattice
using NLCEs and exact diagonalization of finite clusters. We have examined the
thermodynamic properties of the model with exchange parameters appropriate for
\rb, and obtained results that are consistent with the experimental
observations. We have also developed a Lanczos-based NLCE and studied the
approach to the uniform KLHM at $T=0$ in such a pinwheel-dimerized system. We
find strong evidence for a phase transition before the uniform limit is reached,
implying that the KLHM may not spontaneously dimerize in this pattern and its
ground-state phase is stable to small pinwheel distortions.

\section*{Acknowledgments}

This work was supported by the NSF under Grants No.\ OCI-0904597 (E.K. and M.R.)
and No. DMR-1004231 (R.R.P.S.).

\section*{APPENDIX: NUMERICAL LINKED-CLUSTER EXPANSION FOR THE DISTORTED
KAGOME LATTICE}
\label{app:1}

In standard series expansions for the distorted kagome-lattice Heisenberg 
models,~\cite{r_singh_07} the ground-state properties of each cluster in the 
series are calculated using perturbation theory in the small parameter, 
$\lambda$. The contributions of all of the clusters, up to a certain size, 
to each power of $\lambda$ are then added properly into the series. In 
contrast, in the NLCEs, the property of each cluster is computed exactly, i.e.,
to all orders of $\lambda$, using exact diagonalization
techniques.~\cite{m_rigol_06} Because of this, the NLCE does not require
$\lambda$ to be small, and one can achieve convergence in a region that extends
beyond that of the usual series expansions. The convergence in NLCEs is
controlled by the relevant correlation length in the system. Hence, unlike for
standard series expansions, by adding larger cluster sizes, we can always
improve the convergence.

\begin{figure}[!b]
\centerline {\includegraphics*[width=3.3in]{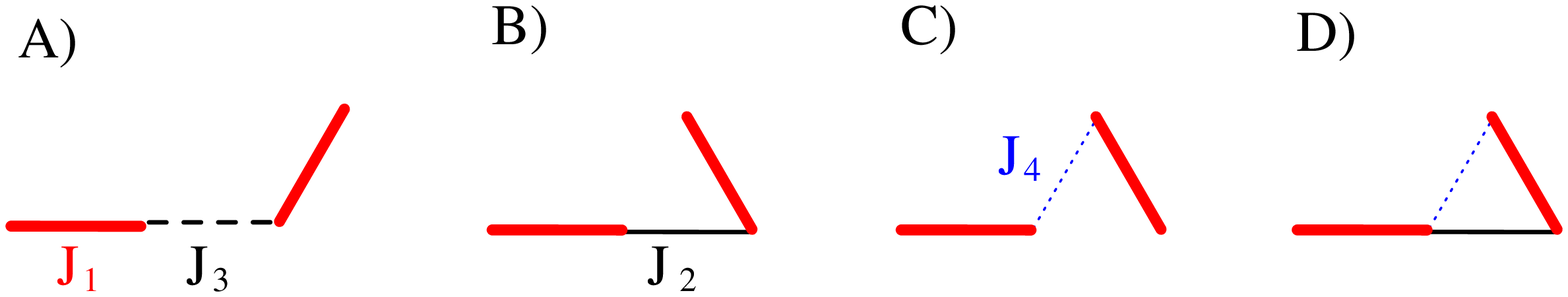}} 
\caption{(Color online) All possible configurations for two neighboring dimers
(red thick bonds) on a kagome lattice with pinwheel anisotropy. In general, four
different types of bonds, $J_1,\dots, J_4$, can be distinguished in this
lattice. Cluster D, in which the two dimers are connected via a complete
triangle, is the only cluster considered in the second order of our expansion.}
\label{fig:second}
\end{figure}

\begin{figure}[!t]
\centerline {\includegraphics*[width=3.3in]{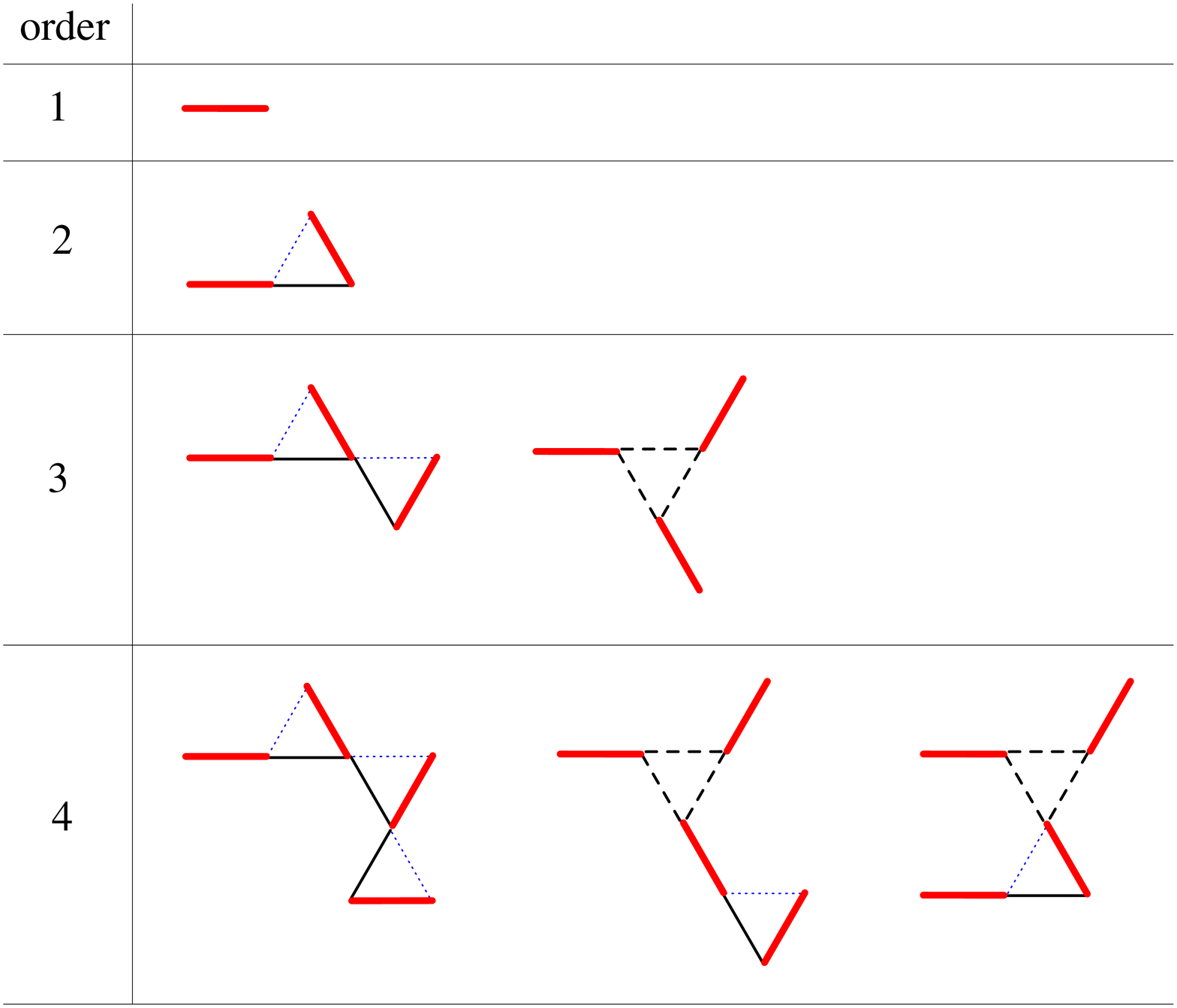}} 
\caption{(Color online) All the topologically distinct clusters up to the fourth
order in our expansion. Bond types are the same as in Fig.~\ref{fig:second}.}
\label{fig:lambda1}
\end{figure}

In our expansion, the building blocks for generating the linked clusters are  
dimers of the pinwheel VBC. This means that in the $n$th order, we generate all
the clusters with $n$ dimers that can be embedded in the anisotropic kagome
lattice (Fig.~\ref{fig:lattice}). Starting from a single dimer in the first
order, there are many possibilities for the topology of the clusters in the
second order, generated by adding another neighboring dimer. However, among 
those, only the four clusters shown in Fig.~\ref{fig:second} are topologically
distinct~\cite{note2}. Next, we require that dimers be connected only through
{\em complete triangles}. For example, in the second order, we consider only
cluster D among those in Fig.~\ref{fig:second}. The above condition
significantly reduces the number of clusters we have to diagonalize in higher
orders while still capturing the fundamental physics at low energies in
relatively short expansions. This selection of the clusters offers a consistent
scheme for the linked-cluster expansion of the infinite lattice. Following the
same process, we end up with two topologically distinct clusters in the third
order by adding the next dimer, and so on. The clusters in the series, up to the
fourth order, are depicted in Fig.~\ref{fig:lambda1}. In Table.~\ref{tb:1}, we
have listed the number of topological clusters in each order up to the
$13^\text{th}$ order.

\begin{table*}[t]
\caption{Number of topologically distinct clusters up to the $13^\text{th}$ order 
of the expansion.}
\begin{tabular}{|l|ccccccccccccc|}
\hline
order number & \ \ 1 \ \ & \ \ 2 \ \ & \ \ 3 \ \ 
& \ \ 4 \ \ & \ \ 5 \ \ & \ \ \ 6 \ \ \ & \ \ \ 7 \ \ \ & \ \ \ 8 \ \ \ & \ \ \ 9 \ \ \ &
 \ \ \ 10 \ \ \ & \ \ \ 11 \ \ \ & \ \ \ 12 \ \ \ & \ \ \ 13 \ \\
\hline
\# of clusters \ & 1 & 1 & 2 & 3 & 7 & 14 & 25 & 50 & 102 & 216 & 442 & 929 & \ 1969 \ \\
\hline
\end{tabular}
\label{tb:1}
\end{table*}

The expansion described above offers a unique approach in which the series
capture not only the high-temperature physics, like the standard NLCEs 
used previously,~\cite{m_rigol_06,e_khatami_11} but also capture the
low-temperature physics very well, as long as the pinwheel VBC is the ground
state. This makes the NLCE used in this work fundamentally different from the
ones implemented before. It uses a basis tailored to describe the ground state
of this particular model. Hence, we use it to calculate properties both at
finite temperatures and, specially, at zero temperature. For finite-temperature
calculations, we compute the full spectrum of the Hamiltonian for every cluster
in the series using standard full diagonalization techniques, and for
zero-temperature calculations, we employ the Lanczos algorithm to access its
ground state only. The advantage of the latter is that it enables us to deal
with systems with much larger Hilbert spaces, and so we are able to carry out
the expansion to significantly larger orders in comparison to the 
finite-temperature case.

Computationally, the dominant part of the calculations is the diagonalization
step in the last order. Ultimately, we are limited either by the large number of
clusters that exist in the largest order we consider and the computational time
it takes to solve for each of them, or by memory requirements for diagonalizing
the Hamiltonian of each of those clusters. For the full diagonalization
finite-temperature calculations, we consider clusters up to the $9$th order, and
for zero-temperature Lanczos calculations, up to the $13$th order. Hence, only
in the last order of the former case, 102 clusters need to be solved. Each of
them has $18$ sites and, therefore, the size of the Hilbert space for the
largest spin sector (with equal number of spin-ups and spin-downs) is
$\binom{18}{9}=48,620$. The large memory requirements for the calculation of the
full spectrum, needed for the finite temperature NLCE, is the one that limits
our expansion to the $9$th order in this case.

For the ground-state calculations, we consider clusters up to the
$13$th order. Here, only in the last order, 1969 clusters have to be
diagonalized. Each of them has 26 sites and the size of the Hilbert space for
the largest spin sector is $\binom{26}{13}=10,400,600$. The memory requirements
for these calculations are much smaller than those for the full diagonalization
(only three vectors of this size need to be stored at any given time). What
limits our ground-state expansion to the $13$th order is the time they take to
converge in the Lanczos loop, so that the ground-state energy is obtained with
a relative accuracy of $10^{-14}$ for orders less than or equal to 12 and of
$10^{-9}$ for the $13$th (last) order. The convergence requirements
that are imposed guarantee at least 5-digit accuracy for the final observables
after summing the contributions of all of the clusters.


\begin{thebibliography}{99}
\bibitem{note2}
\bibinfo{note}{Note that when adding a dimer, we always increase the size of
the cluster by 2; all clusters in the $n$th order have $2n$ sites.}
\end{thebibliography}
\end{document}